\documentclass[%
 reprint,
 nofootinbib,
 amsmath,amssymb,
 aps,
]{revtex4-2}

\usepackage[dvipsnames]{xcolor}
\usepackage[hidelinks]{hyperref}
\hypersetup{
    colorlinks = true,
    linkcolor={Blue}, citecolor={Blue}}
\usepackage{amsmath}
\usepackage{graphicx}% Include figure files
\usepackage{dcolumn}% Align table columns on decimahttps://www.overleaf.com/project/62dd72dbc81c0072b0e16d01l point
\usepackage{bm}% bold math
\begin{document}

\preprint{APS/123-QED}

\title{Interacting Tsallis holographic dark energy in $q$-modified DGP braneworld}% Force line breaks with \\
%\thanks{\texttt{nwl\char`_qcd@yahoo.com}}%

\author{Naouel Boulkaboul}
\email{nwl\textunderscore qcd@yahoo.com}
\affiliation{Independent researcher, Algiers, Algeria}%

\date{\today}% It is always \today, today,

\begin{abstract}
We explore the cosmological aspects of interacting Tsallis holographic dark energy (THDE) in a $q$-modified Dvali–Gabadadze–Porrati (DGP) braneworld setup emerging from non-Gaussian statistical mechanics. To this end, three classes of
superstatistics, that is, log-normal, inverse $\chi^2$ and $\chi^2$ superstatistics were incorporated into the model. We examined the implication of the three superstatistics on different cosmological parameters, namely, the dimensionless energy density and the equation-of-state (EoS) of THDE, along with the deceleration parameter and the squared speed of sound. As a result, we noted that the cosmological parameters stemming from the $\chi^2$ superstatistics, with a parameter $q>1$, represent the highest deviation from those ascribed to the standard DGP model. While the system parameters show appropriate behavior in all three cases, the model cannot achieve stability throughout the history of  the Universe. It is probably the outcome of setting the Hubble horizon as the infrared cutoff. Furthermore, the behavior of EoS was found to be governed by the value of the THDE parameter $\delta$. That is to say, for $\delta>2$ THDE exhibits a phantom-like behavior while for $\delta<2$ it displays a quintessence behavior. Constrained by the dominant energy condition, an upper bound  on $\delta$ ($\delta<2$) has been imposed.
\end{abstract}

\maketitle

%\tableofcontents

\section{\label{sec:level1}Introduction}

 Based on recent astronomical observations, such as distant supernovae Ia’s observation and cosmic microwave background (CMB), it is well known that our Universe is undergoing an accelerated expansion era \cite{riess1998observational,perlmutter1999measurements,de2000flat,knop2003new}. The latter is presumed to be driven by a cosmological constant or vacuum energy \cite{bass2015vacuum}. Meanwhile, due to the absurdly discrepancy between the observed value of the vacuum energy and the values derived from quantum field theory, the cosmological constant problem is the most puzzled conceptual problem in modern cosmology. Looking for alternatives to Einstein’s general relativity in an endeavour to address such a problem, a braneworld model, namely the Dvali-Gabadadze-Porrati (DGP) model has been brought forward \cite{dvali20004d}. In such a setup, one considers our observable Universe to be a (3+1) four-dimensional hypersurface (the brane) embedded in a higher-dimensional spacetime (the bulk). Alongside the Arkani-Hamed-Dimopoulos-Dvali (ADD) \cite{arkani1998hierarchy} and the Randall-Sandrum (RS) \cite{randall1999large} models that were proposed to address the hierarchy problem between the Planck scale and the electroweak scale, the DGP formalism assumes all standard model particles to be confined to the brane, whereas gravity is allowed to propagate along the extra dimension. The astonishing feature of the DGP model stems from the fact that it is a self-accelerating scenario giving rise to an accelerated expansion with vanishing cosmological constant \cite{koyama2008cosmological,xu2014confronting}. Nevertheless, such a self-accelerating scheme is plagued by ghost instability \cite{koyama2007ghosts,gorbunov2006more,charmousis2006dgp,deffayet2006perturbations}.  The Universe acceleration may also be attributed to yet another promising setup, the holographic dark energy (HDE). Numerous efforts in this regard have been established, so far \cite{kim2006equation,del2011holographic,li2004model,liu2010modified,zhang2012generalized,jamil2009generalized,sharma2022interacting,saha2022renyi,saridakis2018ricci,adhikary2021barrow,gong2004extended,bandyopadhyay2011modified,ghaffari2019holographic,nojiri2006unifying,nojiri2017covariant,nojiri2022barrow,nojiri2019modified}. Yet, the primary model which relies on the Bekenstein entropy and the Hubble horizon as its IR cutoff, cannot trigger an accelerated scenario unless an interaction between two dark components of the Universe is accounted for \cite{guberina2005generalized,tavayef2018tsallis,zimdahl2007interacting,hsu2004entropy}. As a remedy, a more generalized HDE, involving quantum gravity considerations \cite{wang2017holographic,wang2016dark}, known as Tsallis holographic dark energy (THDE) has been suggested and was shown to work reasonably well regarding the Universe acceleration \cite{ghaffari2019tsallis,pandey2022new,tavayef2018tsallis,kumar2022quintessence,zadeh2018note,feizi2022interacting,saridakis2018holographic,nojiri2022nonextensive}. Such a connection between thermodynamics and gravity on black hole physics, manifests itself in various aspects as well \cite{padmanabhan2010thermodynamical}, including the entropic
origin of gravity proposed by Verlinde \cite{verlinde2011origin}, which states that gravity can be identified with an entropic force originated from changes in the information linked to the positions of massive bodies, and Padmanabhan's idea that relies on the equipartition argument to provide a thermodynamic interpretation of gravity \cite{padmanabhan2010classical}. Backed up with a new idea which is centered on the claim that the cosmic expansion or equivalently the emergence of space is being driven towards holographic equipartition, two years later Padmanabhan argued that the accelerated expansion of the Universe, along with the standard Friedmann equation arises due to the difference between the surface
degrees of freedom and the bulk degrees of freedom in a region of emerged space \cite{padmanabhan2012emergence}. Such a proposal was applied in the context of equilibrium thermodynamics and
equilibrium statistical mechanics. However, complex gravitational and cosmological systems that are out of equilibrium or exhibit local equilibrium should be dealt with using distributions other than those provided by equilibrium statistical mechanics, i.e. distributions emerging out of nonextensive statistical mechanics \cite{tsallis2009introduction,jahromi2018generalized}, to mention only a few. The latter emerge, as a particular case, from the so-called superstatistics \cite{beck2017superstatistics}, an approach that has proven to be surprisingly prospering in describing nonequilibrium systems that exhibit fluctuations \cite{ourabah2019superstatistics,ourabah2021fingerprints,ourabah2020quasiequilibrium,yalcin2018generalized,beck2017superstatistics,beck2009superstatistics,ayala2018superstatistics,wilk2000interpretation,beck2001dynamical,martinez2022superstatistics}. The backbone of superstatistics reflects the fact that such an apparently complex systems can be regarded as a superposition of two (or more) simple systems operating on different scales. The statistical properties of the system emerge then as a superposition of different statistics.
 
Inspired by the above approaches, we are interested in studying the cosmological consequences of superstatistics on interacting THDE in the DGP braneworld cosmology. Particularly, we will use Padmanabhan’s idea in similar fashion to reference \cite{sheykhi2013emergence} to derive the q-modified Friedmann equation in DGP braneworld, afterwards we will examine the evolution of dark energy density, equation-of-state, and deceleration parameters with respect to the redshift. No less important than the aforementioned parameters, the squared speed of sound will also be investigated in order to obtain clues about the model stability. It might be known that the noninteracting case of THDE in brane cosmology, inculding DGP braneworld, has already been studied in Ref \cite{ghaffari2019tsallis}. The cosmological parameters were investigated for different values of THDE parameter $\delta$. In the spirit of our analysis, though, we restrict ourselves to a fixed value of $\delta$ since our main purpose is to explore superstatistical implications on the model.

 The paper is organized as follows. In section \ref{sec:gi}, following \cite{ourabah2019superstatistics}, we briefly discuss the equipartition theorem stemming from the three main classes of superstatistics. In section \ref{sec:dgp}, we lay out the $q$-modified DGP braneworld scenario associated with the given superstatistics, making use of the corresponding equipartition theorem juxtaposed with Padmanabhan's formalism. We consider a setup in which there is an interaction between dark energy and dark matter. In section \ref{sec:thde}, we discuss the implication of fluctuations on THDE cosmological parameters. In section \ref{sec:stability}, we investigate the stability of the model within the context of superstatistics. In the last section we present our final remarks.\\
\section{\label{sec:gi}Modified equipartition theorem}
In the present section, following
the approach provided in Ref. \cite{ourabah2019superstatistics}, we derive the equipartition law of energy emerging from three different classes of superstatistics, log-normal, inverse $\chi^2$ and $\chi^2$ superstatistics. To begin with, let us consider a thermodynamic system of noninteracting particles, at equilibrium temperature $T$, described by a Maxwell-Boltzmann (MB) velocity distribution,
\begin{equation}
\label{MB velocity}
f_\text{MB}(v)=\sqrt{\frac{\beta m}{2\pi}}\text{exp}\bigg[-\frac{\beta m v^2}{2}\bigg],
\end{equation}
where $\beta\equiv 1/k_B T$ is the inverse temperature in energy units. The temperature presented above is well known to be constant. Yet, complex systems in which the temperature may fluctuate over time should be dealt with using nonequilibrium statistical mechanics where a more generalized distribution is needed. Fluctuations that operate on a long timescale \cite{jizba2018transitions} can be easily incorporated in an equilibrium statistical mechanics taking into account the superposition of different statistics. The nonequlibirm system can be partitioned into small subsystems exhibiting local equilibrium characterized by a sharp value of the inverse temperature $\beta$. As $\beta$ varies adiabatically across the different subsystems, the latter acquire a distribution $f(\beta)$. Consequently, the whole system ends up following a MB velocity distribution, which holds in each subsystem, averaged over $f(\beta)$,
\begin{equation}
\label{superstatics velocity}
B(v)=\int_{0}^{\infty}d\beta f(\beta)\sqrt{\frac{\beta m}{2\pi}}\text{exp}\bigg[-\frac{\beta m v^2}{2}\bigg],
\end{equation}

In short, the approach from which the above distribution is derived is known as superstatistics. Regarding the distribution $f(\beta)$, three fundamental superstatistical classes have to be distinguished \cite{beck2005time}. These three classes consist of:
\paragraph{ Log-normal superstatistics.\textbf{\textemdash}}In
 this case, the inverse temperature $\beta$ follows a log-normal distribution,

\begin{equation}
\label{log-normal}
f(\beta)=\frac{1}{\sqrt{2\pi}s\beta}\text{exp}\bigg({\frac{-(\text{ln}\frac{\beta}{\mu})^2}{2s^2}}\bigg),
\end{equation}
where $\mu$ and $s^2$ are mean and variance parameters.
\paragraph{Inverse-$\chi^2$ superstatistics.\textbf{\textemdash}}
In this case, $\beta$ follows an inverse-$\chi^2$ distribution,
\begin{equation}
\label{inverse chi}
f(\beta)=\frac{\beta_0}{\Gamma\big(\frac{n}{2}\big)}\bigg(\frac{n\beta_0}{2}\bigg)^{n/2}\beta^{-n/2-2}e^{-\frac{n\beta_0}{2\beta}},
\end{equation}
where $\beta_0$ is the average of $\beta$, i.e. $ \beta_0\equiv\int_{0}^{\infty}\beta f(\beta)d\beta$, and $n$ is a real parameter.
\paragraph{$\chi^2$ superstatistics.\textbf{\textemdash}}
 In such a case, $\beta$ is distributed according to the $\chi^2$ distribution of degree $n$,
\begin{equation}
\label{ chi}
f(\beta)=\frac{1}{\Gamma\big(\frac{n}{2}\big)}\bigg(\frac{n}{2\beta_0}\bigg)^{n/2}\beta^{n/2-1}e^{-\frac{n\beta}{2\beta_0}},
\end{equation}

The velocity distribution emerging from the three superstatistics has been derived in Ref \cite{ourabah2019superstatistics}, to which the reader is referred for elaborate considerations, 
\begin{equation}
\label{eq:velocity}
\langle v^2 \rangle_i=\phi_i(q)\langle v^2 \rangle_{\text{MB}} \qquad (i=1, 2, 3),
\end{equation}
where $\langle v^2 \rangle_\text{MB}$ represents the Maxwell-Boltzmann velocity distribution and $\phi_i(q)$ is given by
 \begin{equation}
\label{phi1}
\phi_1(q)\equiv q \qquad (q>0),
\end{equation}
\begin{equation}
\label{phi2}
\phi_2(q)\equiv \frac{2q-1}{q} \qquad (q>1/2),
\end{equation}
\begin{equation}
\label{phi3}
\phi_3(q)\equiv \frac{2}{5-3q} \qquad (0<q<5/3),
\end{equation}
The indices 1, 2, 3 correspond, respectively, to log-normal, inverse $\chi^2$ and $\chi^2$ superstatistics. It is worth noting that for all superstatistics $\langle v^2 \rangle_i$  straightforwardly coincides with the MB case in the limit $q \rightarrow1$.\\
From Eq. \eqref{eq:velocity} the equipartition theorem for the three
different superstatistics is written as
\begin{equation}
\label{equipartition}
E_i=\phi_i(q)\frac{N}{2}T \qquad (i=1, 2, 3),
\end{equation}
where in braneworld cosmology, $N$ will be considered as the bulk degrees of freedom, as it will be shown in the following section. 
\section{\label{sec:dgp}$q$-modified DGP model}
In this section, based on Padmanabhan's space emergence proposal, we derive a $q$-modified DGP braneworld model emerging from non-Gaussian statistical mechanics. To set the scene, we consider a five-dimensional metric including a flat $p=3$ brane, with brane coordinates $x^i$ (i=0, 1, 2, 3) and extra dimension coordinate $y$, in such a way that

\begin{eqnarray}
\label{metric}
ds^2_5=\tilde{g}_{AB}dx^Adx^B=B^2(y)\bigg[-dt^2 \nonumber\\ 
+a^2(t)(\frac{1}{1-Kr^2}dr^2+r^2d\Omega)+dy^2\bigg],
\end{eqnarray}
where $B(y)$\footnote{In DGP braneworld, the warp factor is expressed as $B(y)=e^{\epsilon H|y|}$, with $\epsilon=\pm 1$.} is a real function of the extra dimension $y$, $a(t)$ is the scale factor of the Universe and $K$ stands for a  flat, closed, or open universe with 0, 1, and -1 values, respectively.
Built upon Padmanabhan’s assumptions, the number of degrees of freedom on the spherical
surface of Hubble radius $H^{-1}$ reads \cite{padmanabhan2012emergence}
\begin{equation}
\label{n_sur}
N_\text{sur}=4S=\frac{A}{l^2_p}=\frac{4\pi}{l^2_pH^2},
\end{equation}
where $l_p$ is the Planck length, $A=4\pi H^{-2}$ is the area of the Hubble horizon and $S$ is the entropy which obeys the area law. For the sake of simplicity
we set $k_B=1=c=\hbar$ through the paper. Using Eq. \eqref{equipartition}, the bulk degrees of freedom obey the equipartition law of energy arising out of superstatistics. That is
\begin{equation}
\label{n_bulk}
N_\text{bulk}=\frac{2E_i}{\phi_i(q)T} \qquad (i=1,2,3),
\end{equation}
where the temperature associated with the Hubble horizon is assumed to be the Hawking temperature $T=H/2\pi$, and the energy contained inside the Hubble volume $V=4\pi/3H^3$ is the Komar energy
\begin{equation}
\label{komar}
E_\text{Komar}=-(\rho+3p)V,
\end{equation}
The minus sign is to ensure that $N_\text{bulk}>0$ for an accelerating scenario with $\rho+3p<0$. Here $\rho$ and $p$ include, respectively, the energy densities and pressure on the brane, namely a pressureless matter component with energy density $\rho_m$, sum of baryonic and cold dark matter (CDM) energy densities, and a holographic dark energy component with an energy density $\rho_{\Lambda}$ and a pressure $p_\Lambda$.\\
According to Padmanabhan, the increase $dV$ of the cosmic volume in an infinitesimal interval $dt$ of cosmic time is given by
\begin{equation}
\label{volume derivative}
\frac{dV}{dt}=l^2_p(N_\text{sur}-N_\text{bulk}),
\end{equation}
In a 5-dimensional bulk however, one has to replace $l^2_p$ with $G_5$ which yields
\begin{equation}
\label{bulkvolume derivative}
\frac{d\tilde{V}}{dt}=G_5(N_\text{sur}-N_\text{bulk}),
\end{equation}
with $G_5$ being the gravitational constant on the bulk. The entropy of the apparent horizon
for a 3-dimensional brane embedded in a 5-dimensional Minkowski bulk is given by \cite{sheykhi2007thermodynamical}
\begin{equation}
\label{entropy brane}
S=\frac{\pi H^{-2}}{4G_4}-\epsilon\frac{2\pi H^{-3}}{3G_5},
\end{equation}
where we used $\tilde{r}_A=H^{-1}$ with $\tilde{r}_A$ being the apparent horizon radius. $\epsilon=\pm1$ represents two branches of the DGP model, the $\epsilon=+1$ branch is a self-accelerating solution where the Universe may undergo an accelerating phase with no need of dark energy while the $\epsilon=-1$  (or normal) branch requires a dark energy sector in order to generate an accelerating scenario.\\
It is worth noting that the above expression \eqref{entropy brane} is nothing but a sum of two area terms, the first term which depicts the usual area formula is the contribution of the 4-dimensional gravity on the brane while the second term corresponds to the 5-dimensional area in the bulk. \\
From Eq. \eqref{n_sur}, one can deduce the number of degrees of freedom on the surface to be
\begin{equation}
\label{n brane}
N_\text{sur}=\frac{\pi H^{-2}}{G_4}-\epsilon\frac{8\pi H^{-3}}{G_5},
\end{equation}
If one defines the effective surface as
\begin{equation}
\label{effective surface}
\tilde{A}=4G_5S=\frac{8\pi}{3}\bigg[\frac{3G_5H^{-2}}{2G_4}-\epsilon H^{-3}\bigg],
\end{equation}
the latter varies with respect to time as
\begin{equation}
\label{effective surface derivative}
\frac{d\tilde{A}}{dt}=-8\pi\dot{H}H^{-3}\bigg[\frac{G_5}{G_4}-\epsilon H^{-1}\bigg],
\end{equation}
which leads to an effective volume increase of the form
\begin{equation}
\label{effective volume}
\frac{d\tilde{V}}{dt}=\frac{1}{2H}\frac{d\tilde{A}}{dt}=-4\pi\dot{H}H^{-4}\bigg[\frac{G_5}{G_4}-\epsilon H^{-1}\bigg],
\end{equation}
Matching Eqs. \eqref{effective volume} and \eqref{bulkvolume derivative} and using the equipartition law stemming from superstatistics \eqref{n_bulk}, we end up with 
\begin{equation}
\label{derivative derivative}
\frac{G_5}{2G_4}(2\dot{H}+H^2)-\epsilon H^{-1}(\dot{H}+2H^2)=-\frac{4\pi G_5}{3\phi_i(q)}(\rho+3p),
\end{equation}
which after integration gives rise to
\begin{equation}
\label{derivative derivative}
\frac{G_5}{2G_4}H^2-\epsilon H=\frac{4\pi G_5}{3\phi_i(q)}\rho,
\end{equation}
where we used the continuity equation \footnote{Note that the overall density on the brane satisfies the usual continuity equation, $\dot{\rho}+3H(\rho+p)=0$.} and assumed the constant of integration to be zero. Introducing the crossover length scale between the small and large distances in DGP braneworld as
\begin{equation}
\label{cross over}
r_c=\frac{G_5}{2G_4},
\end{equation}
Eq. \eqref{derivative derivative} can be recast into
\begin{equation}
\label{Friedmann DGP}
\phi_i(q)[H^2-\frac{\epsilon}{r_c} H]=\frac{\rho}{3M^2_p},
\end{equation}
which is our $q$-modified Friedmann equation in DGP braneworld, with $M^2_p=1/8\pi G_4$ being the usual reduced Planck mass. One may bear in mind that for $r_c>>r$ the standard 4-dimensional Einstein's theory of gravity is recovered.\\
Now, Assuming an interacting dark energy scenario, one can introduce, for an interaction of the form $Q=\Gamma{\rho}_{\Lambda}$, the energy densities' continuity equations as

\begin{equation}
\label{continutity equation1}
\dot{\rho}_{\Lambda}+3H(1+\omega_{\Lambda})\rho_{\Lambda}=-Q,
\end{equation}
\begin{equation}
\label{continutity equation2}
\dot{\rho}_\emph{m}+3H\rho_\emph{m}=Q.
\end{equation}
This is nothing more than a decaying mechanism in which the holographic dark energy sector decays into CDM at a decaying rate $\Gamma$. The total conservation of the energy-momentum tensor is ensured by the equality of Q. Note that we have introduced the equation-of-state (EoS) parameter $\omega_{\Lambda}=\frac{P_{\Lambda}}{\rho_{\Lambda}}$, where $p_{\Lambda}$ is the pressure associated with the holographic component. Given the fact that the ratio of the two energy densities has the form $r=\rho_m/\rho_{\Lambda}$, the above equations can be rewritten as
\begin{equation}
\label{continutity equation1'}
\dot{\rho}_{\Lambda}+3H(1+{\omega}^{\text{eff}}_{\Lambda})\rho_{\Lambda}=0,
\end{equation}
\begin{equation}
\label{continutity equation2'}
\dot{\rho}_\emph{m}+3H(1+{\omega}^{\text{eff}}_{m})\rho_\emph{m}=0.
\end{equation}
with ${\omega}^{\text{eff}}_{\Lambda}={\omega}_{\Lambda}+\frac{\Gamma}{3H}$ and ${\omega}^{\text{eff}}_{m}=-\frac{\Gamma}{3Hr}$, being the effective EoS parameters.
The time evolution of the ratio $r$ is then given by
\begin{equation}
\label{ratio derivative}
\dot{r}=3Hr\bigg[\omega_{\Lambda}+\frac{1+r}{r}\frac{\Gamma}{3H}\bigg]=3Hr\big[{\omega}^{\text{eff}}_{\Lambda}-{\omega}^{\text{eff}}_{m}\big]
\end{equation}
In the present work we use the notation $\Gamma=3b^2(1+r)H$ with $b^2$ being the coupling constant. Nevertheless, in our analysis we will take into account both cases ($b^2 =0$ and $b^2\neq0$).\\\\
Now, let us define the dimensionless energy density parameters as
\begin{equation}
\label{reduced density}
\Omega_{m}=\frac{\rho_m}{3{M}^2_{p}H^2_0}, \qquad \Omega_{\Lambda}=\frac{\rho_\Lambda}{3{M}^2_{p}H^2_0}, \qquad \Omega_{r_c}=\frac{1}{4{r}^2_{c}H^2_0}
\end{equation}
which, for each superstatistical case, satisfy the normalization condition
\begin{equation}
\label{converted Freidmann equation}
\Omega_m+\Omega_{\Lambda}+2\phi_i(q)\epsilon\sqrt{\Omega_{r_c}}\frac{H}{H_0}=\phi_i(q)\frac{H^2}{H^2_0} \qquad (i=1,2,3)
\end{equation}
Consequently, the density $\Omega_m$ along with $\Omega_{\Lambda}$, as it will be shown in the following section,  will be a function of the nonextensive parameter $q$. Note that the standard normalization condition is recovered for $q=1$, and hence the usual $\Omega_m$ and $\Omega_{\Lambda}$. Setting $H(z)=E(z)H_0$, where $H_0$ is the current value of the Hubble parameter, one can notice the dependence of the ratio $r$ on the parameter $q$ through the normalization condition Eq. \eqref{converted Freidmann equation}.
\begin{equation}
\label{converted ratio}
r=\frac{\phi_i(q)E^2(z)-\Omega_{\Lambda}-2\phi_i(q)\epsilon\sqrt{\Omega_{r_c}}E(z)}{\Omega_{\Lambda}},
\end{equation}
which consequently leads to a relation between $\Omega_{\Lambda}$ and $r$ of the form 
\begin{equation}
\label{omega and r}
\Omega_{\Lambda}=\phi_i(q)\frac{E^2(z)-2\epsilon\sqrt{\Omega_{r_c}}E(z)}{1+r} \qquad (i=1, 2, 3),
\end{equation}
In the following, THDE will be our ultimate candidate for the interacting dark energy scenario. The system  cosmological parameters due to superstatistics will therefore be explored.
\section{\label{sec:thde}Cosmological behavior of THDE in $q$-modified DGP braneworld}
The holographic dark energy density is derived from the entropy-area relation of black holes, $S\propto A$, with $A=4\pi L^2$ being the area of the horizon. However, if quantum gravitational corrections are included, the horizon entropy of a black hole may be modified. Accordingly, a number of modified entropy expressions have been introduced \cite{wang2017holographic,wang2016dark,jahromi2018generalized}, including that suggested by Tsallis
and Cirto as \cite{tsallis2013black}
\begin{equation}
\label{tsallis entropy}
S_T=\gamma A^\delta
\end{equation}
where $A\propto L^2$ is the area of a $d$-dimensional system with characteristic length $L$. $\gamma$ is an unknown constant and $\delta=d/(d-1)$ for $d>1$. Note that the above equation reduces to the additive Bekenstein entropy for $\delta=1$ and $\gamma=2\pi {M}^2_p$.
Within the context of the holographic principle, the degrees of freedom of a given physical system can be projected onto its boundary surface \cite{hooft1993dimensional}. Following such an assumption, and having in hand the relation between the system's entropy, together with the infrared cutoff $L$ and the ultraviolet cutoff $\Lambda$

\begin{equation}
\label{entrpy cutoff}
L^3\Lambda^3\leq S^{3/4}
\end{equation}
we obtain, using Eq. \eqref{tsallis entropy}, the inequality

\begin{equation}
\label{inequality}
\Lambda^4 \leq (4\pi)^\delta \gamma L^{2\delta-4}
\end{equation}
where $\Lambda^4$ represents the THDE energy. It follows that the THDE energy density reads
\begin{equation}
\label{thde}
\rho_\Lambda=BL^{2\delta-4}
\end{equation}
where $B$ is an unknown parameter. It should be noted that for the special case $\delta=2$ the standard cosmological constant case, i.e. $\rho_\Lambda=const$ is restored.
It is worth mentioning that, within the frame of a HDE model, the largest length $L$ of the model must be specified. To this end, we choose the Hubble radius as IR cutoff, i.e. $L=H^{-1}$. Hence, using Eq. \eqref{thde} along with Eq. \eqref{reduced density} , we obtain
\begin{equation}
\label{omegaT}
\Omega_\Lambda=\frac{BH^{4-2\delta}}{3{M}^2_pH^2_0}=\frac{B}{3{M}^2_p}E^{4-2\delta}(z){H^{2(1-\delta)}_0}
\end{equation}
Furthermore, the derivative of $\rho_\Lambda$ is given by
\begin{equation}
\label{thde derivative}
\dot{\rho}_\Lambda=2(2-\delta)\rho_\Lambda \frac{\dot{H}}{H},
\end{equation}
while the derivative of $\Omega_\Lambda$ with respect to $x=$ ln $a$ is expressed as
\begin{equation}
\label{omegaT derivative}
{\Omega'_\Lambda}=2(2-\delta)\Omega_\Lambda \frac{\dot{H}}{H^2}=2(2-\delta)\Omega_\Lambda \frac{E'(z)}{E(z)}
\end{equation}
where we used $\dot{\Omega}_\Lambda={\Omega'_\Lambda}H$ and $\dot{E}(z)={E'(z)}H$.
%Taking the time derivatives of Eq.\ref{thde} and Eq.\ref{omega} respectively, we find
%\begin{equation}
%\label{omega derivative}
%{\Omega}'_\Lambda=2\Omega_\Lambda(1-\delta)\frac{\dot{H}}{H^2}
%\end{equation}
%where prime denotes the derivative respect to $x=${ln} $a$, and we used the relation $\dot{\Omega}_\Lambda=H{\Omega}'_\Lambda$.\\
 Now, combining the time derivative of  Eq. \eqref{Friedmann DGP} with Eqs. \eqref{continutity equation1}, \eqref{continutity equation2}, \eqref{reduced density}, \eqref{thde derivative} and using $\Gamma=3b^2(1+r)H_0E(z)$ along with Eq. \eqref{omega and r}, we get
 \begin{widetext}
\begin{equation}
\label{derivative E1}
E'(z)=\frac{3E(z)\big[\phi_i(q)(E^2(z)-2\epsilon\sqrt{\Omega_{r_c}}E(z))(b^2-1)+\frac{B}{3{M}^2_p}E^{4-2\delta}(z)H^{2(1-\delta)}_0\big]}{2\big[\phi_i(q)(E^2(z)-\epsilon\sqrt{\Omega_{r_c}}E(z))-(2-\delta)\frac{B}{3{M}^2_p}E^{4-2\delta}(z)H^{2(1-\delta)}_0\big]} \qquad (i=1, 2, 3),
\end{equation}
 \end{widetext}
where the prime is the derivative with respect to $x=$ln $a$. Expressing Eq. \eqref{omegaT} at $z=0$, with $E(z=0)=1$ and $\Omega_{\Lambda }(z=0)=\Omega_{\Lambda 0}$ represent the present epoch, the above equation can be rewritten as
  \begin{widetext}
\begin{equation}
\label{derivative E2}
E'(z)=\frac{3E(z)\big[\phi_i(q)(E^2(z)-2\epsilon\sqrt{\Omega_{r_c}}E(z))(b^2-1)+E^{4-2\delta}(z)\Omega_{\Lambda 0}\big]}{2\big[\phi_i(q)(E^2(z)-\epsilon\sqrt{\Omega_{r_c}}E(z))-(2-\delta)E^{4-2\delta}(z)\Omega_{\Lambda 0}\big]} \qquad (i=1, 2, 3),
\end{equation}
 \end{widetext}
and from which one can get the derivative of  $\Omega_\Lambda$ with respect to ln $a$ as
  \begin{widetext}
 
\begin{equation}
\label{derivative ln omega}
{\Omega'_\Lambda}=(2-\delta){\Omega_\Lambda}\frac{3\big[\phi_i(q)(E^2(z)-2\epsilon\sqrt{\Omega_{r_c}}E(z))(b^2-1)+\Omega_{\Lambda}\big]}{\phi_i(q)(E^2(z)-\epsilon\sqrt{\Omega_{r_c}}E(z))-(2-\delta)\Omega_{\Lambda}} \qquad (i=1, 2, 3),
\end{equation}
  \end{widetext}
Yet, it would be more feasible to express ${\Omega'_\Lambda}$ as a function of ${\Omega_\Lambda}$ instead of $E(z)$. Hence, since we have ${\Omega_\Lambda}=\tilde{\Omega}_\Lambda E^2(z)$ with $\tilde{\Omega}{_{\Lambda}}=\frac{\rho_\Lambda}{3M^2_pH^2}$, one ends up with the following equation
 \begin{widetext}
\begin{equation}
\label{derivative ln omega tilde}
\tilde{\Omega}'_\Lambda=(1-\delta){\tilde{\Omega}_\Lambda}\frac{3\big[\phi_i(q)(1-2\epsilon\sqrt{\Omega_{r_c}}\tilde{\Omega}{_\Lambda}^{-\frac{1}{2(1-\delta)}}\Omega{_{\Lambda 0}}^{\frac{1}{2(1-\delta)}})(b^2-1)+\tilde{\Omega}{_\Lambda}\big]}{\phi_i(q)(1-\epsilon\sqrt{\Omega_{r_c}}\tilde{\Omega}{_\Lambda}^{-\frac{1}{2(1-\delta)}}\Omega{_{\Lambda 0}}^{\frac{1}{2(1-\delta)}})-(2-\delta)\tilde{\Omega}_\Lambda} \qquad (i=1, 2, 3),
\end{equation}
  \end{widetext}
Interestingly, the above equation is reduced to the standard equation of THDE in DGP brane model in the limit $b=0$ and $q\rightarrow 1$. Furthermore, for $r_c>>1$ (or equivalently $\Omega_{r_c}\rightarrow0$) where the effects of the extra dimension are unnoticed, THDE in the 4-dimensional Einstein gravity is recovered.

\begin{figure*}

\includegraphics[scale=0.45]{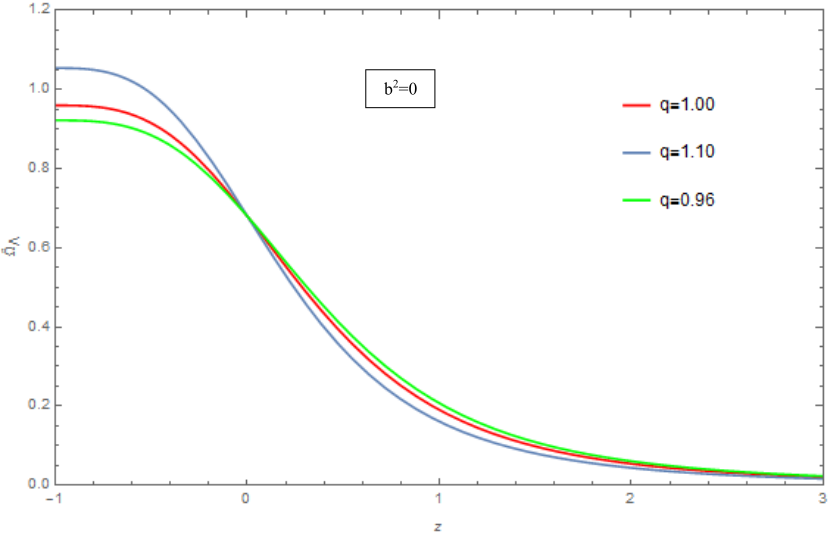}% Here is how to import EPS art
\includegraphics[scale=0.45]{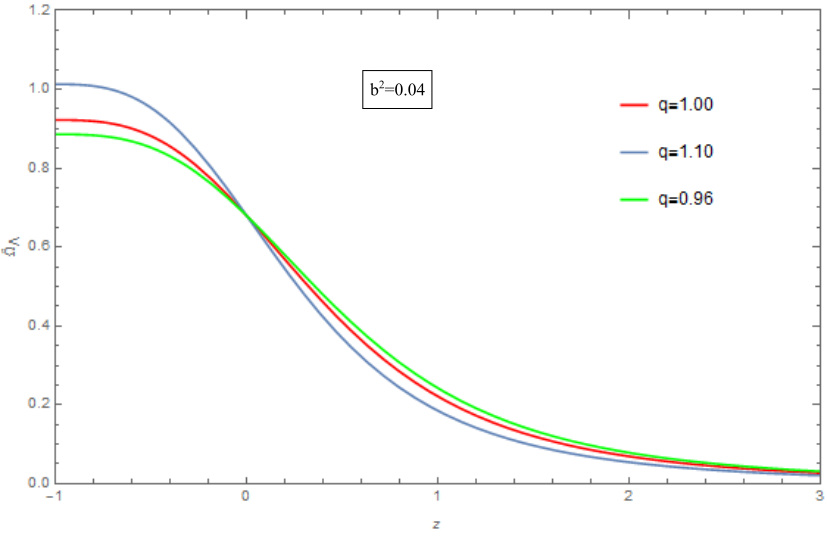}
\textbf{a. log-normal superstatistics $\phi_1(q)\equiv q$   $(q>0)$}\\

\includegraphics[scale=0.45]{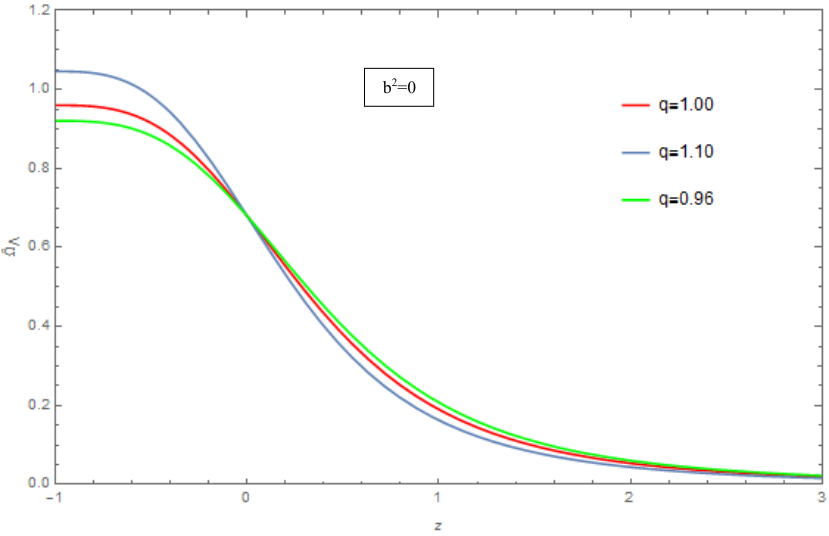}% Here is how to import EPS art
\includegraphics[scale=0.45]{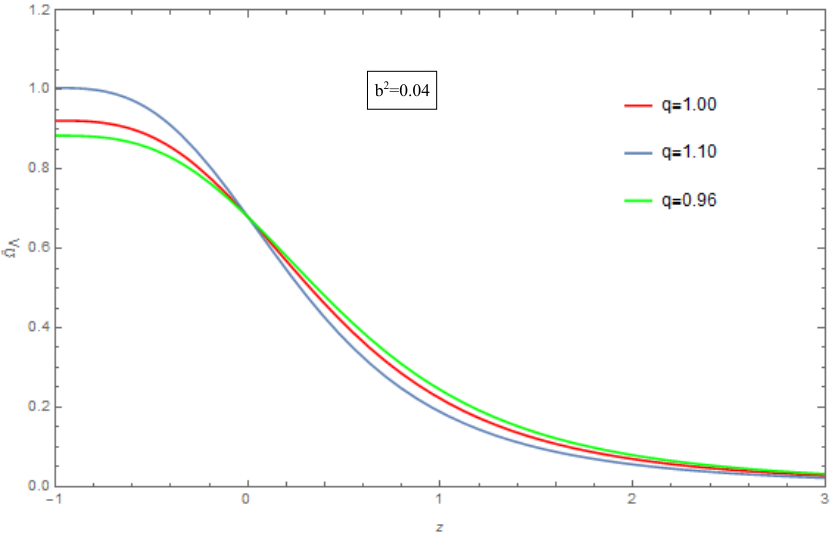}
\textbf{b. Inverse $\chi^2$ superstatistics $\phi_2(q)\equiv \frac{2q-1}{q}$   $(q>1/2)$}
\includegraphics[scale=0.45]{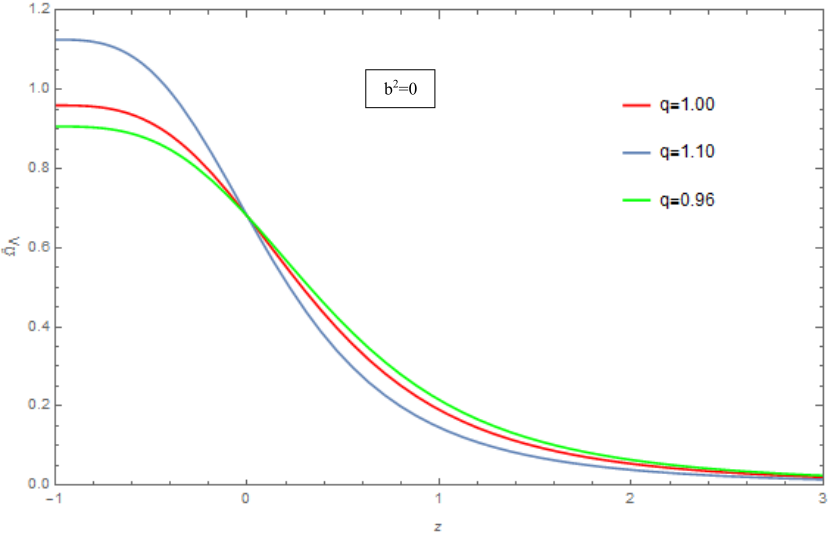}% Here is how to import EPS art
\includegraphics[scale=0.45]{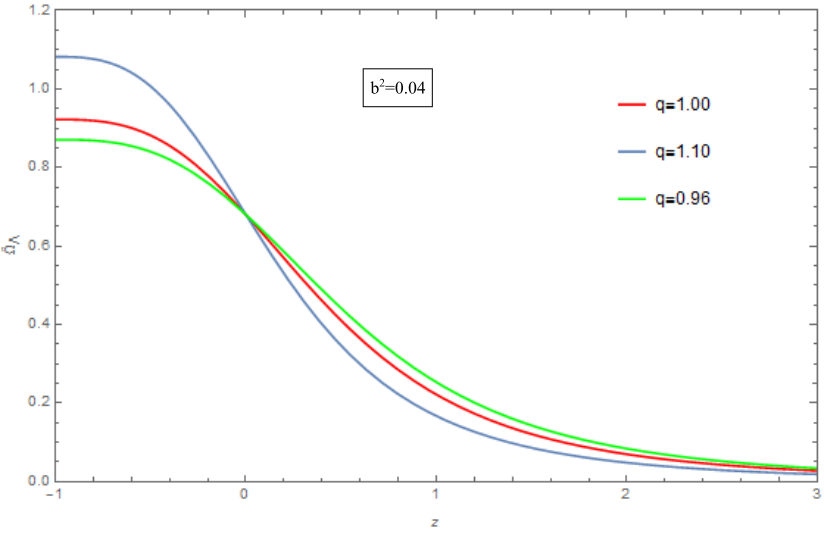}
\textbf{c. $\chi^2$ superstatistics $\phi_3(q)\equiv \frac{2}{5-3q}$   $(0<q<5/3)$}
\caption{\label{fig:omega}The evolution of $\tilde{\Omega}_\Lambda$ versus $z$ for $\tilde{\Omega}_{\Lambda 0}=0.68$, $\Omega_{r_c}=0.0003$, $\epsilon=1$,  $\delta=2.2$, $b^2=0$ (left), $b^2=0.04$ (right) and some values of $q$.}

\end{figure*}

\begin{figure*}
\includegraphics[scale=0.45]{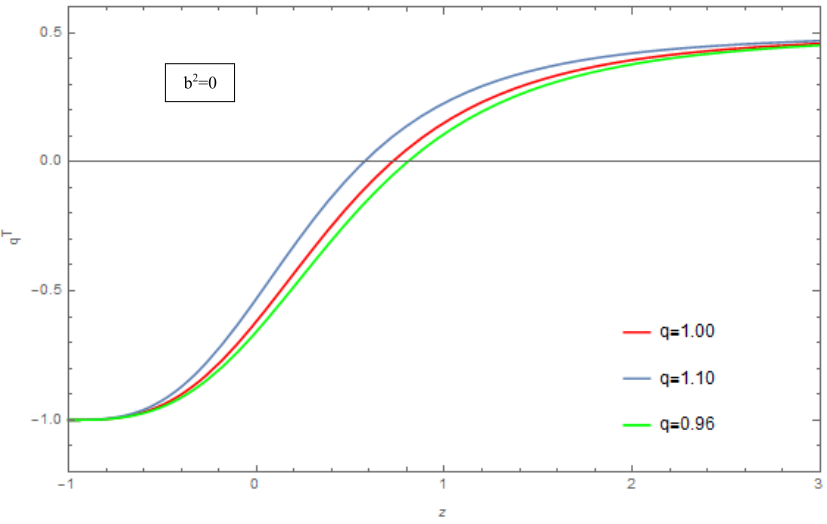}% Here is how to import EPS art
\includegraphics[scale=0.45]{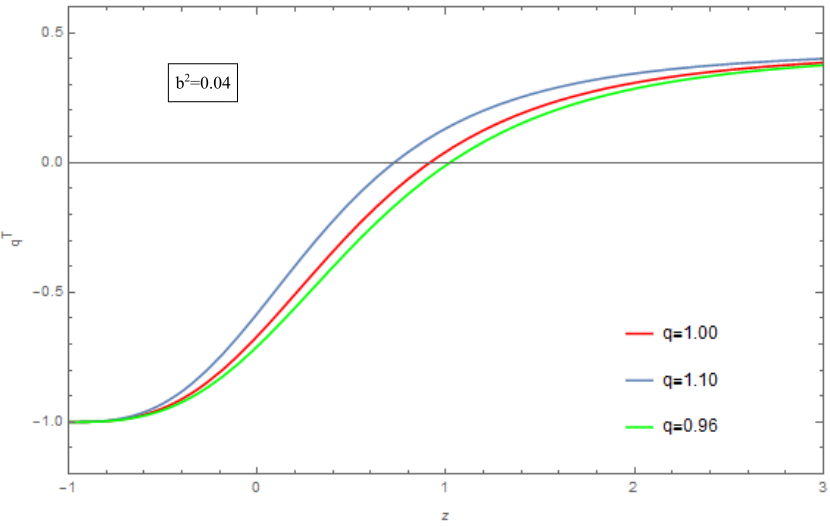}
\textbf{a. log-normal superstatistics $\phi_1(q)\equiv q$   $(q>0)$}
\includegraphics[scale=0.45]{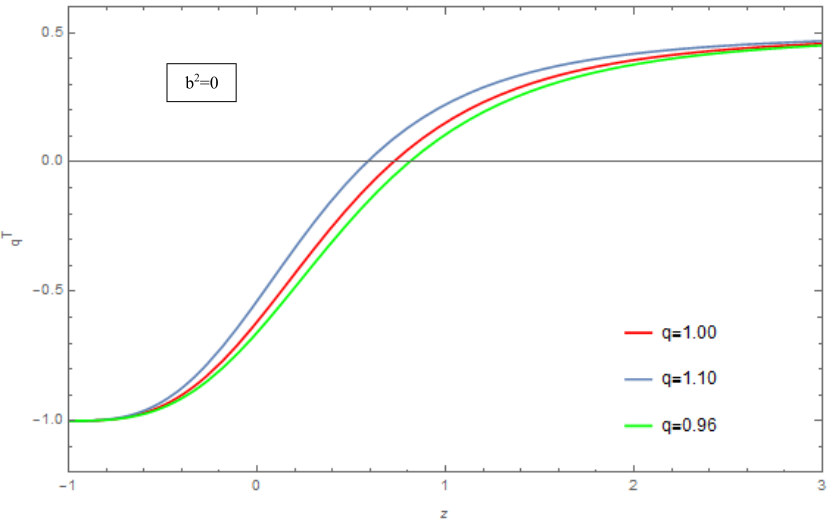}% Here is how to import EPS art
\includegraphics[scale=0.45]{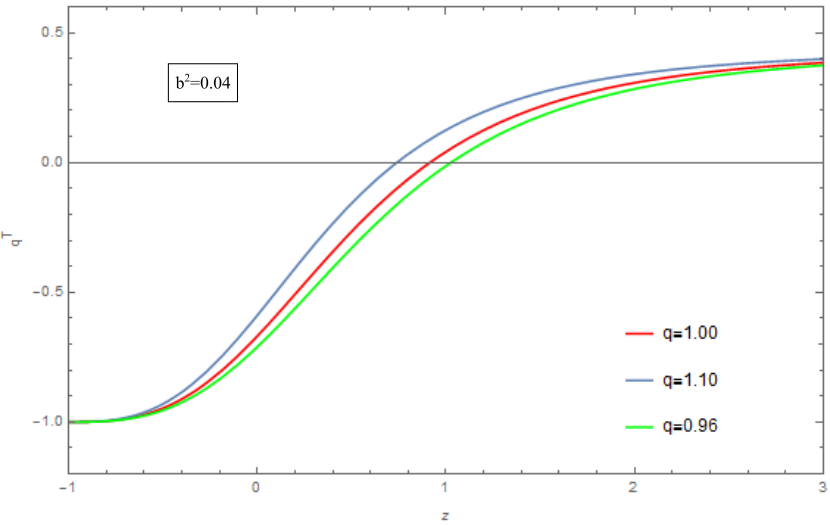}
\textbf{b. Inverse $\chi^2$ superstatistics $\phi_2(q)\equiv \frac{2q-1}{q}$   $(q>1/2)$}
\includegraphics[scale=0.45]{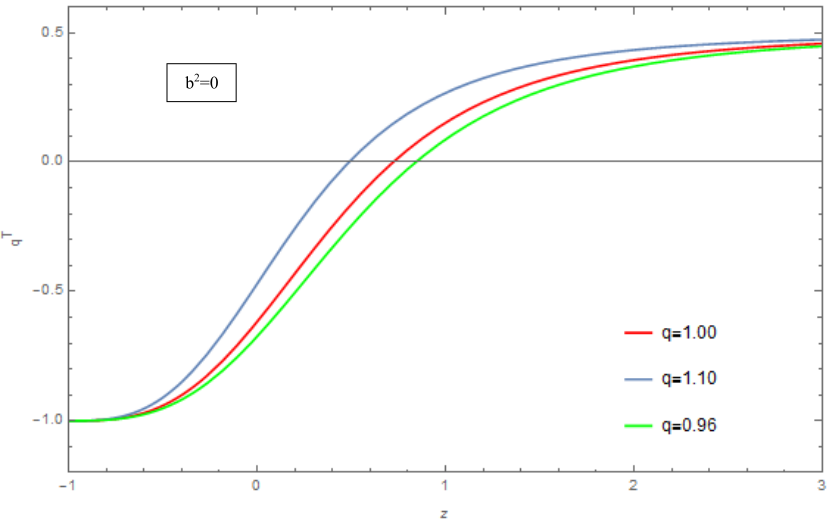}% Here is how to import EPS art
\includegraphics[scale=0.45]{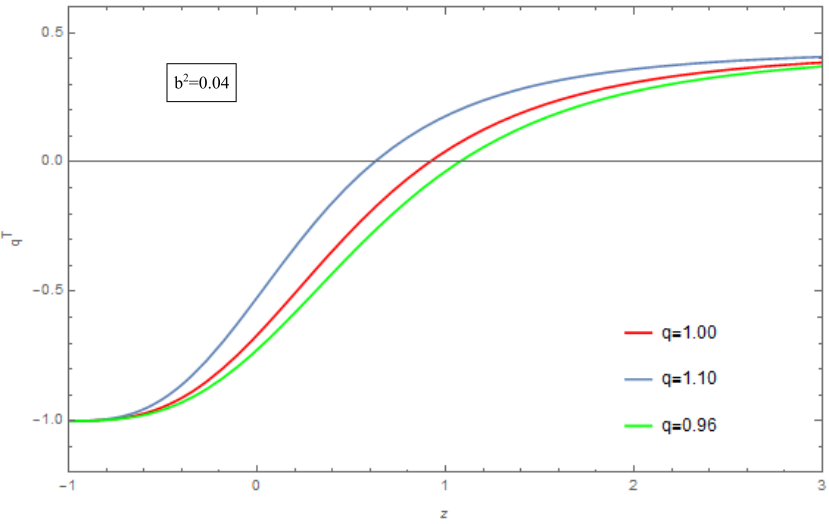}
\textbf{c. $\chi^2$ superstatistics $\phi_3(q)\equiv \frac{2}{5-3q}$   $(0<q<5/3)$}
\caption{\label{fig:q}The evolution of $q^T$ versus $z$ for $\tilde{\Omega}_{\Lambda 0}=0.68$, $\Omega_{r_c}=0.0003$, $\epsilon=1$,  $\delta=2.2$, $b^2=0$ (left), $b^2=0.04$ (right) and some values of $q$.}
\end{figure*}

\begin{figure*}
\includegraphics[scale=0.45]{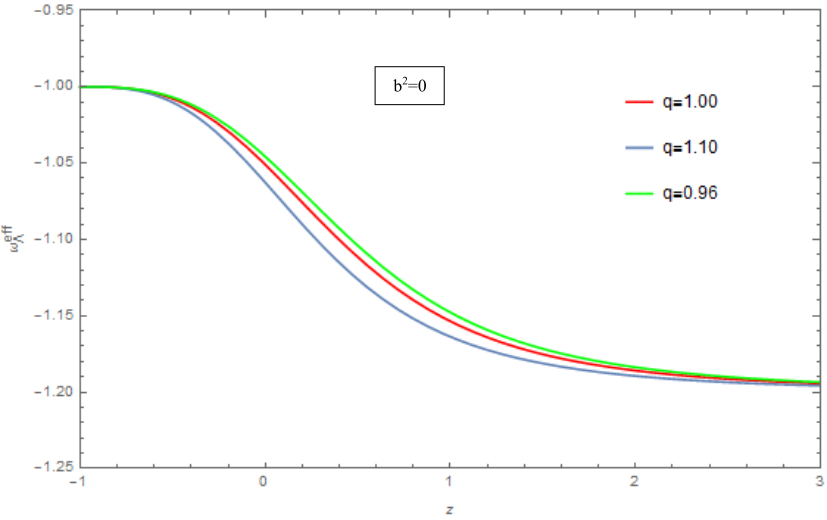}% Here is how to import EPS art
\includegraphics[scale=0.45]{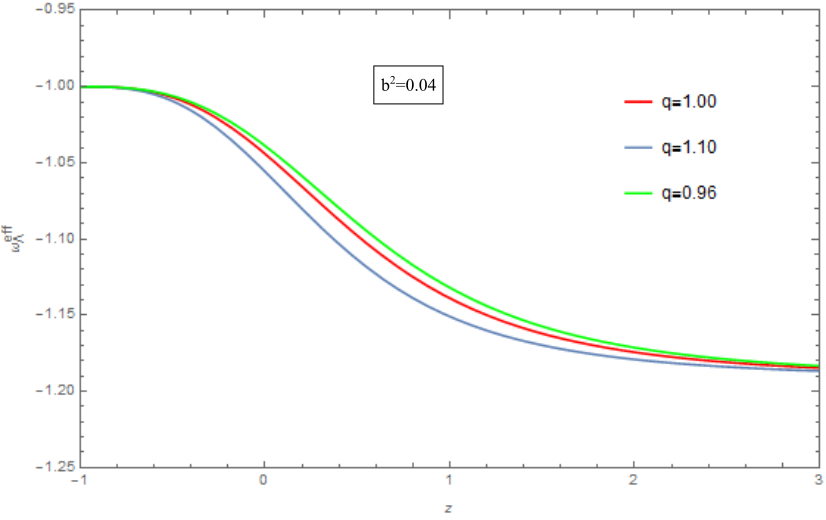}
\textbf{a. log-normal superstatistics $\phi_1(q)\equiv q$   $(q>0)$}
\includegraphics[scale=0.45]{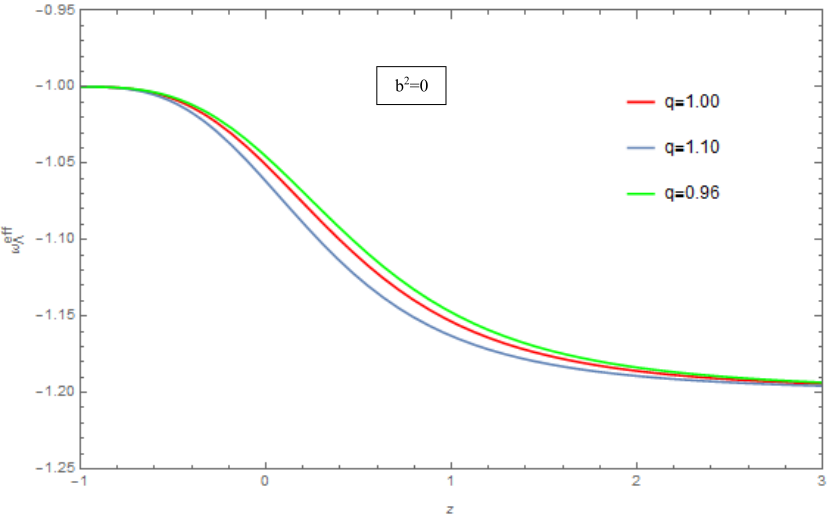}% Here is how to import EPS art
\includegraphics[scale=0.45]{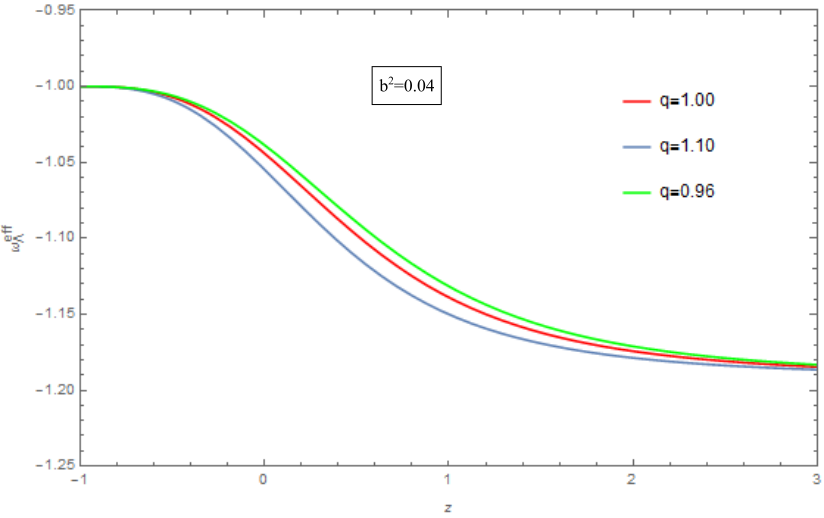}
\textbf{b. Inverse $\chi^2$ superstatistics $\phi_2(q)\equiv \frac{2q-1}{q}$   $(q>1/2)$}
\includegraphics[scale=0.45]{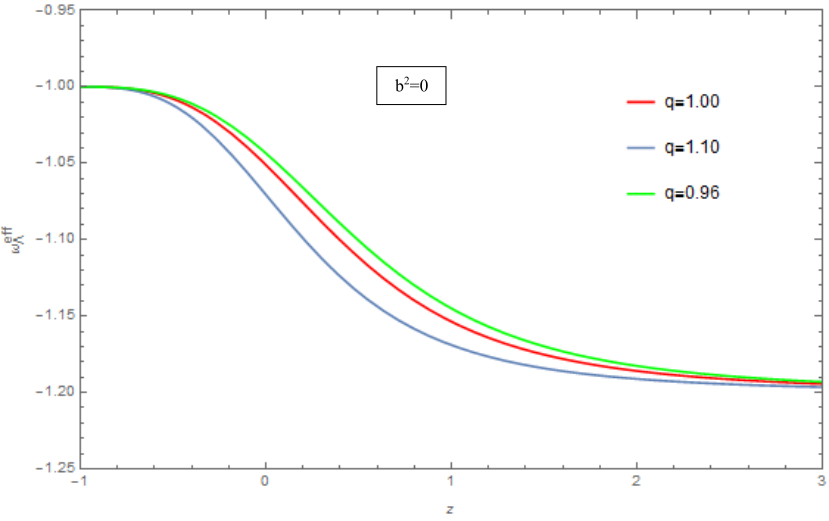}% Here is how to import EPS art
\includegraphics[scale=0.45]{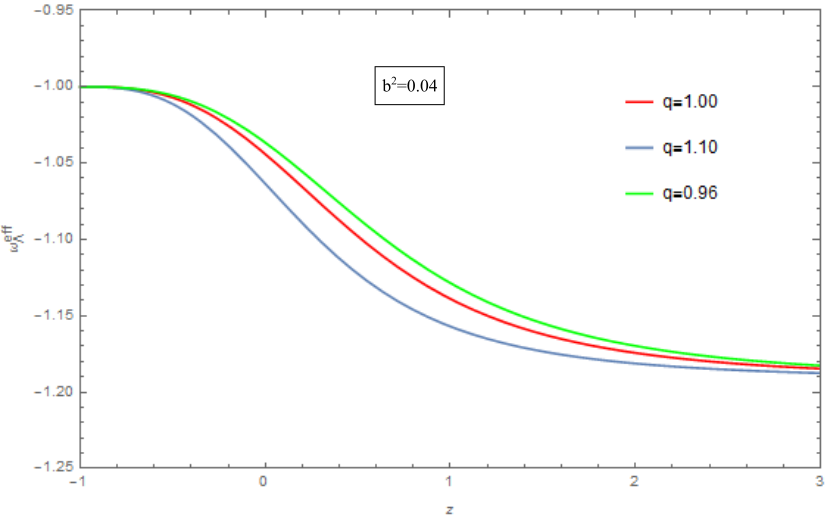}
\textbf{c. $\chi^2$ superstatistics $\phi_3(q)\equiv \frac{2}{5-3q}$   $(0<q<5/3)$}
\caption{\label{fig:EoS}The evolution of the EoS parameter ${\omega^{\text{eff}}_\Lambda}$  versus $z$ for $\tilde{\Omega}_{\Lambda 0}=0.68$, $\Omega_{r_c}=0.0003$, $\epsilon=1$,  $\delta=2.2$, $b^2=0$ (left), $b^2=0.04$ (right) and some values of $q$.}
\end{figure*}

To investigate the impact of superstatistics on the dimensionless energy density $\tilde{\Omega}_\Lambda$, the deceleration $q^T$\footnote{The superscript $T$, which refers to Tsallis, is to distinguish the deceleration parameter from the nonextensive parameter $q$.} and the effective EoS ${\omega^{\text{eff}}_\Lambda}$, we inspect both cases $q\geq1$ ($q=1$ and $q=1.1$) and $q<1$ ($q=0.96$).\\
Setting $\delta=2.2$, $\epsilon=1$, $\Omega{_{\Lambda 0}}=\tilde{\Omega}_\Lambda(z=0)=0.68$ and $\Omega_{r_c}=0.0003$, the variation of $\tilde{\Omega}_\Lambda$ with respect to redshift $z$ is shown in Fig. \ref{fig:omega}.  The non-interacting scenario is shown on the left hand side while the interacting one, with a coupling $b^2=0.04$, is illustrated on the right hand side. It is easy to check that for $z\rightarrow\infty$,  $\tilde{\Omega}_\Lambda\rightarrow0$ which is in agreement with the fact that the matter component was dominant at early time. One may also notice that the effect of the nonextensive parameter $q$ is barely noticed for $q=0.96$, peculiarly at past time ($0<z<\infty$). A slight difference, in the variation of $\tilde{\Omega}_\Lambda$ with respect to $z$, is spotted for $q=1.1$. The difference is more pronounced in the case of $\chi^2$ superstatistics, particularly in the range $-1<z<0$ (future). This finding is corroborated by the fact that the $\chi^2$ superstatistics represents the higher deviation of the standard Maxwell-Boltzmann statistics when $q$ moves away from 1 as has been shown in Ref. \cite{ourabah2019superstatistics}.
From Fig. \ref{fig:omega} it is apparent that the transition from matter to dark energy domination is affected by the fluctuations. In the future era, dark energy dominates over matter ($\tilde{\Omega}_\Lambda(z\rightarrow -1)\rightarrow \phi_i$ ($i=1, 2, 3$)) for the noninteracting case of  $b^2 = 0$, while for the interaction case, dark energy decays into CDM till the two components become comparable due to energy transfer. Moreover, for all three superstatistics, the dimensionless energy density $\tilde{\Omega}_\Lambda$ decreases as $q$ increases for a redshift lying in the range $0<z<\infty$ (early Universe) while it increases as $q$ increases for a redshift lying within the range $-1<z<0$ (late Universe).\\\\
%The effective EoS parameters are then given by
%\begin{equation}
%\label{new effective EoS parameter1}
%{\omega}^{eff}_\Lambda=-1+(2-\delta)\frac{(\phi_i(q)-\Omega_\Lambda-2\phi_i(q)\epsilon\sqrt{\Omega_{r_c}})-b^2(\phi_i(q)-2\phi_i(q)\epsilon\sqrt{\Omega_{r_c}})}{\phi_i(q)(1-\epsilon\sqrt{\Omega_{r_c}}-(2-\delta)\Omega_\Lambda)}
%\end{equation}
%\begin{equation}
%\label{new effective EoS parameter2}
%{\omega}^{eff}_m=\frac{b^2(\phi_i(q)-2\phi_i(q)\epsilon\sqrt{\Omega_{r_c}})}{\phi_i(q)-\Omega_\Lambda-2\phi_i(q)\epsilon\sqrt{\Omega_{r_c}}}
%\end{equation}
Proceeding now to the deceleration parameter $q^T$. The latter is displayed in terms of $\tilde{\Omega}_\Lambda$ as
 \begin{widetext}
\begin{eqnarray}
\label{tsallis deceleration}
q^{T}=-1-\frac{E'(z)}{E(z)} 
 =-1-\frac{3\big[\phi_i(q)(1-2\epsilon\sqrt{\Omega_{r_c}}\tilde{\Omega}_\Lambda^{-\frac{1}{2(1-\delta)}}\Omega_{\Lambda 0}^{\frac{1}{2(1-\delta)}})(b^2-1)+\tilde{\Omega}_\Lambda\big]}{2\big[\phi_i(q)(1-\epsilon\sqrt{\Omega_{r_c}}\tilde{\Omega}_\Lambda^{-\frac{1}{2(1-\delta)}}\Omega_{{\Lambda 0}}^{\frac{1}{2(1-\delta)}})-(2-\delta)\tilde{\Omega}_\Lambda\big]}  \qquad (i=1, 2, 3),
\end{eqnarray}
 \end{widetext}
It should be noted that $q^T>0$ represents a decelerating phase of the Universe, while $q^T< 0$ showcases
the Universe in its accelerating phase. It is worth pointing out that, according to observational data, the transition redshift $z_t$ from a decelerating to an accelerating era, which occurs at $q^T=0$, lies within the range of $0.4<z<1$.\\
Accounting for the three superstatistical cases mentioned above, the evolution of the deceleration parameter $q^T$ versus the redshift $z$, for some values of the nonextensive parameter $q$, is plotted in Fig. \ref{fig:q}. Again, we used the values $\delta=2.2$, $\Omega{_{\Lambda 0}}=\tilde{\Omega}_\Lambda(z=0)=0.68$, $\Omega_{r_c}=0.0003$ and $\epsilon=1$. For all superstatistics, it is obvious that at $z\rightarrow-1$ ($z\rightarrow \infty$), $q^T$ unveils a desired asymptotic behavior, i.e. $q^T\rightarrow-1$ ($q^T\rightarrow\frac{1-3b^2}{2}$) regardless the value of $\delta$. While the transition redshift $z_t$ is shifted towards larger $z$ due to the interaction, the three superstatistics tend to shift $z_t$ towards smaller values as $q$ increases. At this point, it can be checked that, for the given $b^2$ coupling and $q=0.96$, $z_t$ lies outside the suitable range, i.e. $z_t\sim1.03$, $z_t\sim1.04$ and $z_t\sim1.08$ for log-normal, $\chi^2$ inverse and $\chi^2$ superstatistics, respectively. Therefore, an apt mixing between the parameters $q$ and $b^2$ is required in order to ensure an adequate range of transition redshift $z_t$. This can be done by constraining the parameters with observational data and which will be thoroughly explored in a forthcoming paper.\\
On the other hand, the effective EoS parameters are derived using Eqs. \eqref{continutity equation1'}, \eqref{continutity equation2'}
 \begin{widetext}
\begin{equation}
\label{eos omega tsallis}
{\omega^{\text{eff}}_\Lambda}=-1-\frac{2(2-\delta)}{3}\frac{E'(z)}{E(z)}=-1-(2-\delta)\frac{\phi_i(q)(1-2\epsilon\sqrt{\Omega_{r_c}}\tilde{\Omega}_\Lambda^{-\frac{1}{2(1-\delta)}}\Omega{_{\Lambda 0}}^{\frac{1}{2(1-\delta)}})(b^2-1)+\tilde{\Omega}_\Lambda}{\phi_i(q)(1-\epsilon\sqrt{\Omega_{r_c}}\tilde{\Omega}_\Lambda^{-\frac{1}{2(1-\delta)}}\Omega_{\Lambda 0}^{\frac{1}{2(1-\delta)}})-(2-\delta)\tilde{\Omega}_\Lambda},
\end{equation}
 
\begin{equation}
\label{eos m tsallis}
{\omega^{\text{eff}}_m}=-\frac{b^2\phi_i(q)[1-2\epsilon\sqrt{\Omega_{r_c}}\tilde{\Omega}_\Lambda^{-\frac{1}{2(1-\delta)}}\Omega_{\Lambda 0}^{\frac{1}{2(1-\delta)}}]}{\phi_i(q)(1-2\epsilon\sqrt{\Omega_{r_c}}\tilde{\Omega}_\Lambda^{-\frac{1}{2(1-\delta)}}\Omega_{\Lambda 0}^{\frac{1}{2(1-\delta)}})-\tilde{\Omega}_\Lambda} \qquad (i=1, 2, 3),
\end{equation}
 \end{widetext}
Negativeness of the effective EoS of CDM is ascribed to the decaying process. Indeed, CDM production is simultaneously triggered by virtue of the effective nonequilibrium pressure (${P^{\text{eff}}_m}=-\frac{\Gamma\rho_\Lambda}{3H}<0$) via the cosmic anti-frictional force \cite{kim2006equation}. As expected, in the limiting case $q=1$ and the interaction-free case $b^2=0$, the relations derived in Ref. \cite{ghaffari2019tsallis}  are deduced.\\
For each superstatistical case, Fig. \ref{fig:EoS} depicts the evolution of the EoS parameter $\omega^{\text{eff}}_\Lambda$ as a function of redshift $z$ for different values of the nonextensive parameter $q$ with and without an interaction term. One can bear in mind that for an interacting dark energy, the true equation-of-state is given by $\omega^{\text{eff}}_\Lambda$ rather than $\omega_\Lambda$. For the different superstatistical cases with (no) interaction, it is obvious that THDE exhibits a phantom behavior, i.e. $\omega^{\text{eff}}_\Lambda<-1$, at $z\rightarrow0$ and $z\rightarrow \infty$, while $\omega^{\text{eff}}_\Lambda \rightarrow -1$  at $z\rightarrow -1$, revealing that at late time the model mimics the standard cosmological constant behavior. It is worth stressing that the phantom-like behavior is due to the chosen value of $\delta$ $(\delta>2)$. For $\delta<2$, however, $\omega^{\text{eff}}_\Lambda$ will display a quintessence behavior, i.e. $\omega^{\text{eff}}_\Lambda>-1$ (see for instance Eq. \eqref{eos omega tsallis}). According to previously published computations, the phantom-like equation-of-state for the holographic dark energy is forbidden \cite{kim2006equation, d2019holographic}. This condition may not hold true in the THDE case as long as $\delta >2$. Thus, inasmuch as one considers the compatibility of the holographic principle with the dominant energy condition $\rho_\Lambda \geq {|p_\Lambda|}$ we can then impose an upper bound on $\delta$, namely $\delta <2$. The latter is in accordance with the observational constraints performed in Ref. \cite{d2019holographic}. Apart from THDE, a similar feature has been encountered in the Barrow holographic dark energy case where the latter can be quintessence or phantom, or even display a phantom-divide crossing during the cosmological evolution \cite{adhikary2021barrow}.  Regarding the nonextensive parameter on the other side, the value of $\omega^{\text{eff}}_\Lambda$ in the $q$-DGP braneworld scenario for $q>1$ is lower than that in the conventional DGP model as $z\rightarrow0$. Still, little to no distinction is observed at past and future times when $z\rightarrow\infty$ and $z\rightarrow-1$, respectively. At the other extreme, the $b^2$ coupling tends to increase $\omega^{\text{eff}}_\Lambda$ with respect to the interaction-free case.\\

%\onecolumngrid
\section{\label{sec:stability}Stability}
\begin{figure*}
\includegraphics[scale=0.35]{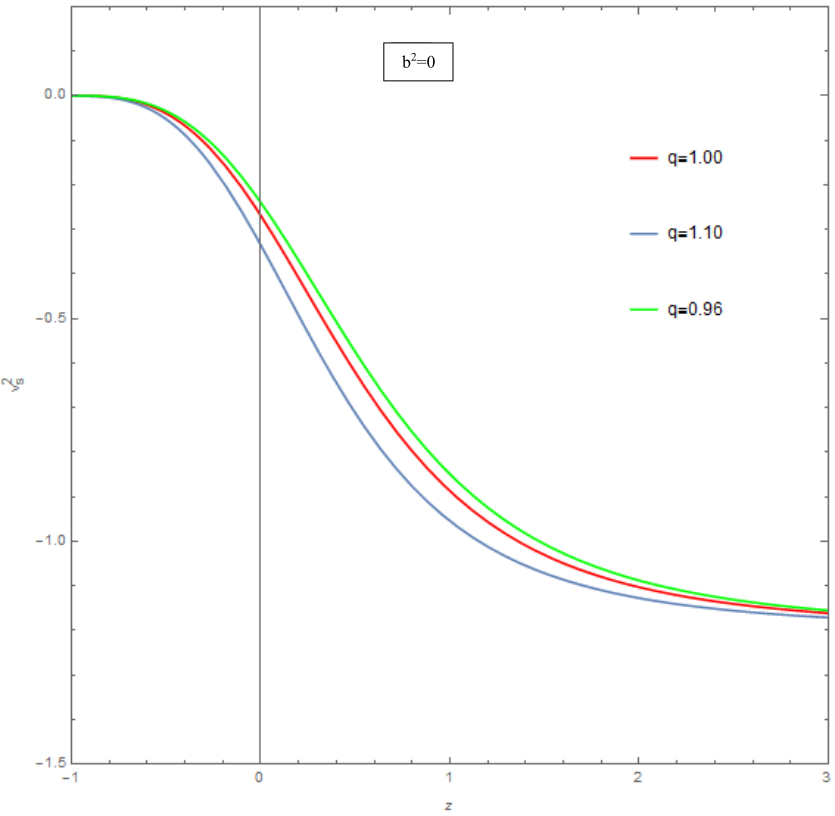}% Here is how to import EPS art\\
\includegraphics[scale=0.35]{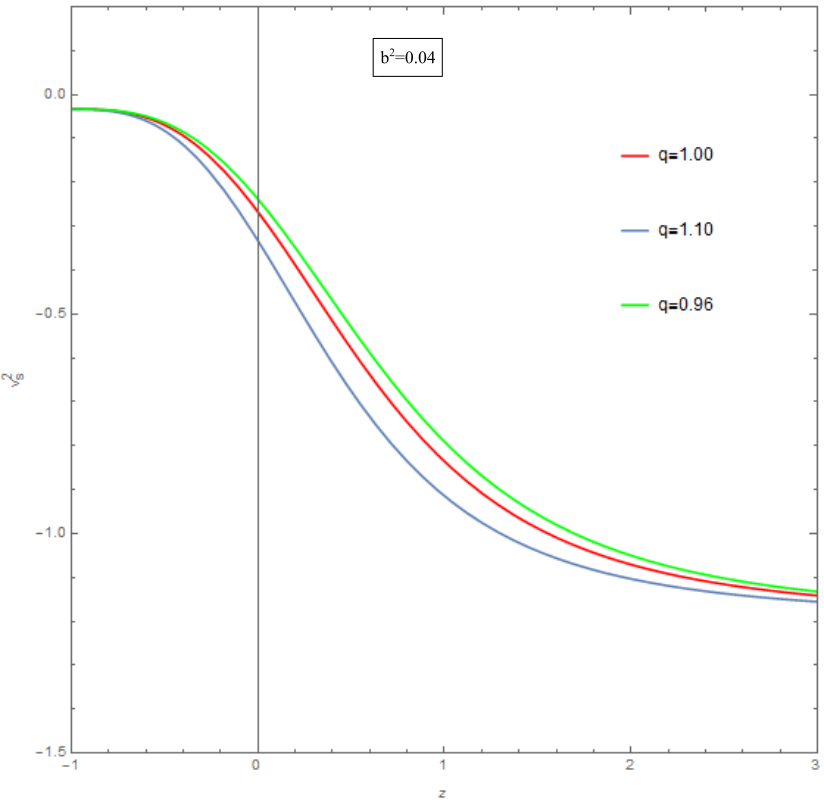}\\
\textbf{a. log-normal superstatistics $\phi_1(q)\equiv q$   $(q>0)$}
\includegraphics[scale=0.35]{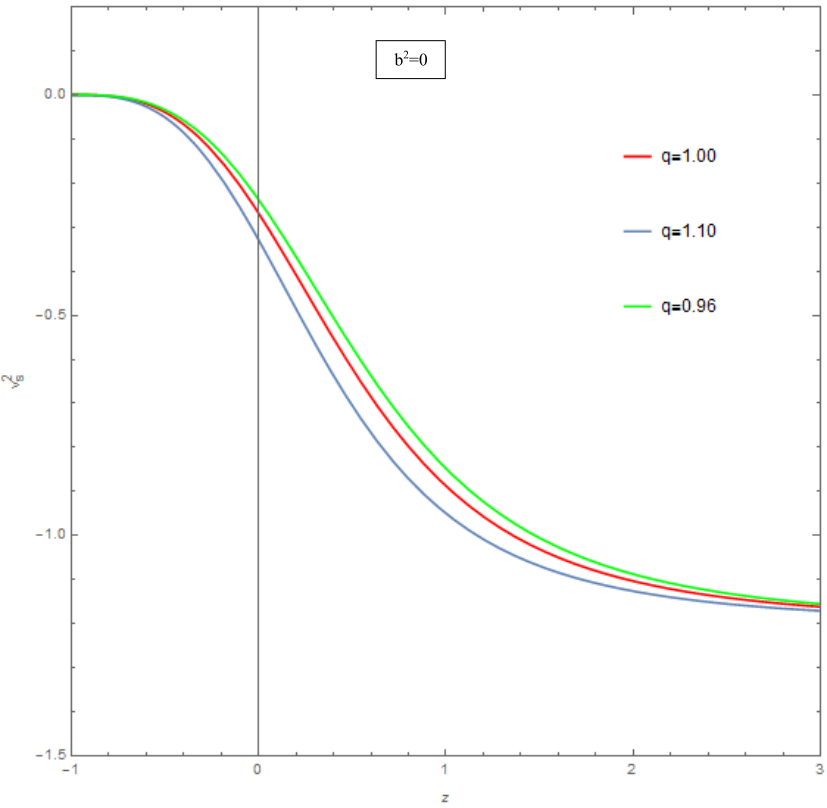}% Here is how to import EPS art
\includegraphics[scale=0.35]{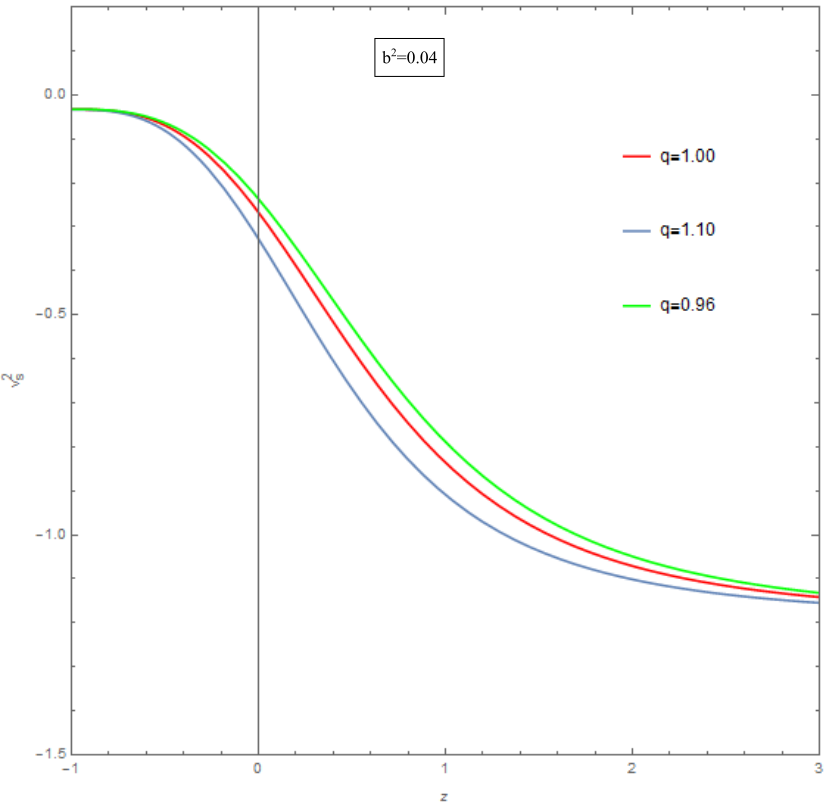}\\
\textbf{b. Inverse $\chi^2$ superstatistics $\phi_2(q)\equiv \frac{2q-1}{q}$   $(q>1/2)$}
\includegraphics[scale=0.35]{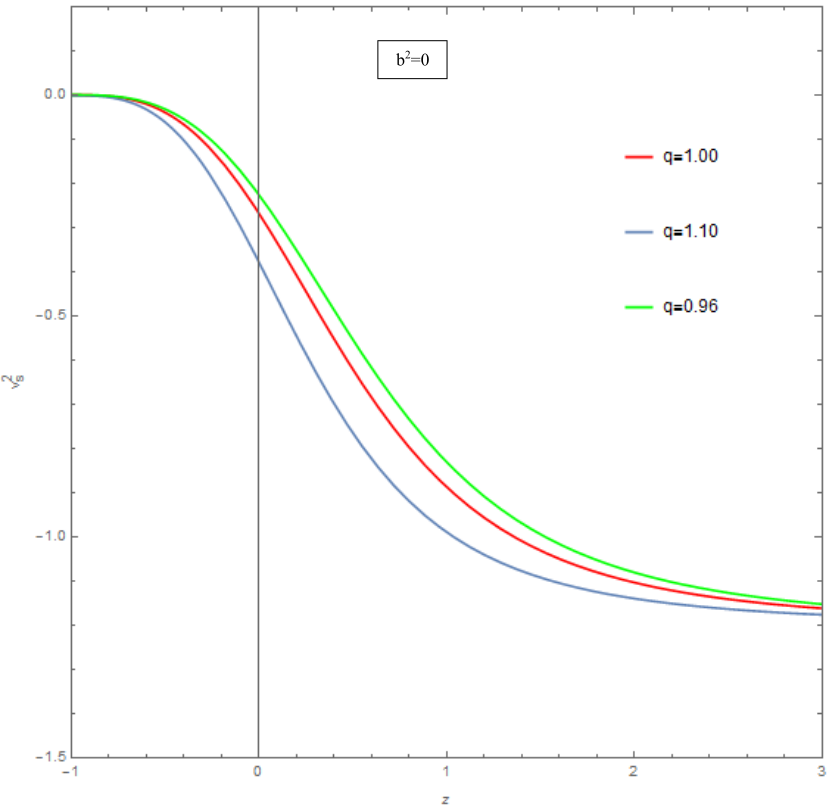}% Here is how to import EPS art
\includegraphics[scale=0.35]{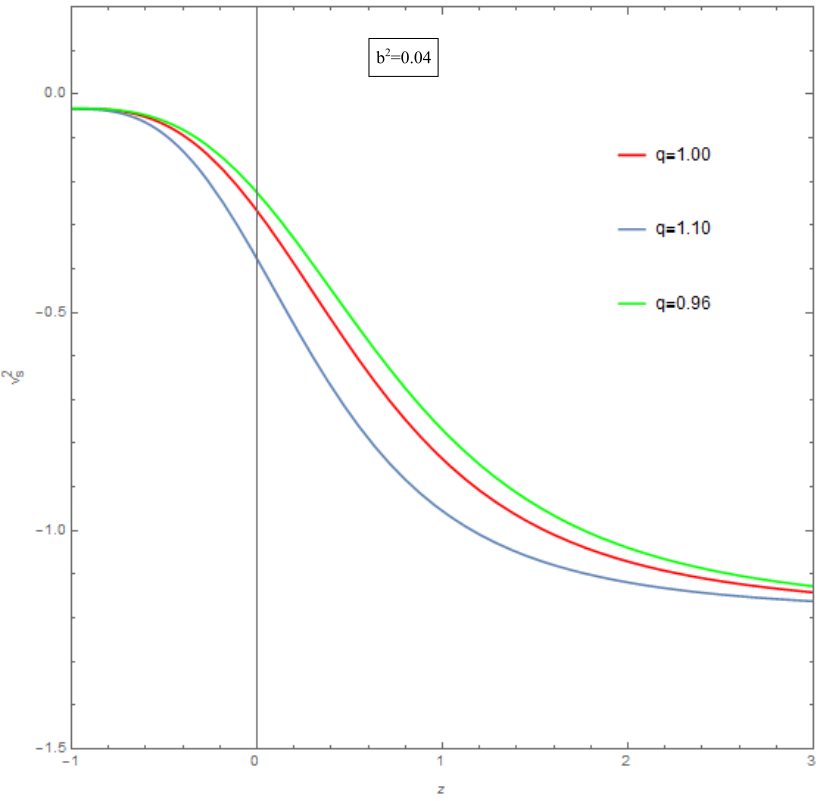}\\
\textbf{c. $\chi^2$ superstatistics $\phi_3(q)\equiv \frac{2}{5-3q}$   $(0<q<5/3)$}
\caption{\label{fig:vs}The evolution of $v^2_s$ versus $z$ for $\tilde{\Omega}_{\Lambda 0}=0.68$, $\Omega_{r_c}=0.0003$, $\epsilon=1$,  $\delta=2.2$, $b^2=0$ (left), $b^2=0.04$ (right) and some values of $q$.}
\end{figure*}

The stability of models against small perturbations is scrutinized through the use of the squared
of the sound speed ${v}^2_s$. The model is said to be stable against perturbations when ${v}^2_s>0$. The squared sound speed ${v}^2_s$ is written as
\begin{equation}
\label{sound speed1}
v^2_s=\frac{dp^{\text{eff}}_\Lambda}{d\rho_\Lambda}= \frac{\dot{p}^{\text{eff}}_\Lambda}{\dot{\rho}_\Lambda},
\end{equation}
where ${p}^{\text{eff}}_\Lambda={\omega}^{\text{eff}}_\Lambda\rho_\Lambda$. Hence, the above equation reduces to
\begin{equation}
\label{sound speed2}
v^2_s={\omega}^{\text{eff}}_\Lambda+\frac{\dot{\omega}^\text{eff}_\Lambda \rho_\Lambda}{\dot{\rho}_\Lambda},
\end{equation}
which, by plugging Eqs. \eqref{continutity equation1'} and \eqref{eos omega tsallis}, yields
\begin{widetext}
\begin{equation}
\label{sound speed3}
v^2_s=-1-(2-\delta)\frac{\phi_i(q)(1-2\epsilon\sqrt{\Omega_{r_c}}\tilde{\Omega}_\Lambda^{-\frac{1}{2(1-\delta)}}\Omega_{\Lambda 0}^{\frac{1}{2(1-\delta)}})(b^2-1)+\tilde{\Omega}_\Lambda}{\phi_i(q)(1-\epsilon\sqrt{\Omega_{r_c}}\tilde{\Omega}_\Lambda{-\frac{1}{2(1-\delta)}}\Omega_{\Lambda 0}^{\frac{1}{2(1-\delta)}})-(2-\delta)\tilde{\Omega}_\Lambda}\nonumber
\end{equation}
\begin{equation}
+\phi_i(q)\tilde{\Omega}_\Lambda\frac{\epsilon\sqrt{\Omega_{r_c}}\tilde{\Omega}_\Lambda^{-\frac{1}{2(1-\delta)}}\Omega_{\Lambda 0}^{\frac{1}{2(1-\delta)}}[(b^2-1)((2-\delta)(3-2\delta)-\frac{\phi_i(q)}{2}\tilde{\Omega}^{-1}_\Lambda)+\frac{3}{2}-\delta]-(1-\delta)((b^2-1)(2-\delta)+1)}{[\phi_i(q)(1-\epsilon\sqrt{\Omega_{r_c}}\tilde{\Omega}_\Lambda^{-\frac{1}{2(1-\delta)}}\Omega_{\Lambda 0}^{\frac{1}{2(1-\delta)}})-(2-\delta)\tilde{\Omega}_\Lambda]^2},
\end{equation}
\end{widetext}
Inspecting Fig. \ref{fig:vs} which illustrates the evolution of $v^2_s$ as a function of redshift $z$ for some values of $q$, it is evident that, for all values of redshift ($z\in [-1, \infty[$), the current model is unstable against perturbations irrespective of the value of $b^2$ or $q$, showing that neither the interaction nor the nonextensive aspects of spacetime have  an apparent effect on THDE stability. Yet, the stability may be sensitive to $\delta$ parameter and/or IR cutoff \cite{saha2020interacting}. Although it has been argued in Ref. \cite{saha2020interacting} that THDE correspondence of interacting Generalized Chaplygin Gas (GCG) model in Kaluza-Klein configuration can reach stability for $\delta<0.63$, this limiting value is unphysical since it leads to negative dimension $d$ \footnote{Note that $\delta$ is related to the system dimension $d$ $(d>1)$ through $\delta=d/(d-1)$.}.
\section{\label{sec:remark}Final remarks}

In all, we sifted the cosmological dynamics of interacting dark energy in a $q$-modified DGP braneworld emerging from superstatistics. By regarding the cosmic dark sector as Tsallis holographic dark energy, we explored the THDE different parameters. While there are numerous cutoffs to choose from, we set the Hubble horizon as our IR cutoff. Results due to log-normal and inverse $\chi^2$ superstatistics are almost indistinguishable from one another. Conversely, $\chi^2$ superstatistics represents perceptible deviations, particularly when $q$ moves away from 1.\\ The transition from a decelerating to an accelerating era relies heavily upon the $q$ parameter and $b^2$ coupling. In contrast, THDE stability is insensitive to such parameters. Roughly speaking, it has been shown that the model stability cannot be reached in the entire history of the Universe. The stability of noninteracting THDE, with Hubble cutoff, is not achieved in 4D gravity either \cite{tavayef2018tsallis}. This drawback may be dealt with considering a different IR cutoff. \\
One of the striking features of THDE is the fact that it can possess a phantom or quintessence behavior depending on the value of the $\delta$ parameter. To the best of our knowledge, this has yet to be mentioned in the literature for a Hubble horizon cutoff. Imposed by the dominant energy condition, values of $\delta>2$ which hinder THDE from being quintessence must be excluded. In the case of future event horizon cutoff, a different bound has been set up in Ref. \cite{saridakis2018holographic}.\\
Although it would be appealing to confront the
current scenario with observational data in order to constrain the parameters $\delta$, $q$, and $b^2$, such investigations are beyond the scope of the present paper and are better left for an upcoming work.
%\twocolumngrid

\nocite{*}
\bibliography{aapmsamp}% Produces the bibliography via BibTeX.

\end{document}